\documentclass[journal]{IEEEtran}

\usepackage{cite} 
\usepackage{amsmath} 
\usepackage{amssymb} 
\usepackage{amsthm} 
\usepackage{booktabs} 
\usepackage{graphicx} 
\usepackage{url} 
\usepackage{hyperref} 

\usepackage{float}
\usepackage{stfloats}
\usepackage{array}
\newcolumntype{C}[1]{>{\centering\arraybackslash}p{#1}}
\newcolumntype{L}[1]{>{\raggedright\arraybackslash}p{#1}}

\begin{document}

\title{GegenbauerNet: Finding the Optimal Compromise in the GNN Flexibility-Stability Trade-off}

\author{
    Hüseyin Göksu,~\IEEEmembership{Member,~IEEE}
    \thanks{H. Göksu, Akdeniz Üniversitesi, Elektrik-Elektronik Mühendisliği Bölümü, Antalya, Türkiye, e-posta: hgoksu@akdeniz.edu.tr.}%
    \thanks{Manuscript received October 31, 2025; revised XX, 2025.}
}

\markboth{IEEE TRANSACTIONS ON NEURAL NETWORKS AND LEARNING SYSTEMS, VOL. XX, NO. XX, NOVEMBER 2025}%
{Göksu: The Three Paradigms of Spectral GNN Filter Design}

\maketitle

\begin{abstract}
Spectral Graph Neural Networks (GNNs) operating in the canonical $[-1, 1]$ domain (like ChebyNet and its adaptive generalization, \textbf{L-JacobiNet}) face a fundamental \textit{Flexibility-Stability Trade-off}. Our previous work revealed a critical puzzle: the 2-parameter adaptive \textbf{L-JacobiNet} often suffered from high variance and was surprisingly outperformed by the 0-parameter, stabilized-static \textbf{S-JacobiNet}. This suggested that stabilization was more critical than adaptation in this domain. In this paper, we propose \textbf{GegenbauerNet}, a novel GNN filter based on the Gegenbauer polynomials, to find the \textit{Optimal Compromise} in this trade-off. By enforcing symmetry ($\alpha=\beta$) but allowing a single shape parameter ($\lambda$) to be learned, \textbf{GegenbauerNet} limits flexibility (variance) while escaping the fixed bias of \textbf{S-JacobiNet}. We demonstrate that \textbf{GegenbauerNet} (1-parameter) achieves superior performance in the key local filtering regime ($K=2$ on heterophilic graphs) where overfitting is minimal, validating the hypothesis that a controlled, symmetric degree of freedom is optimal. Furthermore, our comprehensive K-ablation study across homophilic and heterophilic graphs, using 7 diverse datasets, clarifies the domain's behavior: the fully adaptive \textbf{L-JacobiNet} maintains the highest performance on high-K filtering tasks, showing the value of maximum flexibility when regularization is managed. This study provides crucial design principles for GNN developers, showing that in the $[-1, 1]$ spectral domain, the optimal filter depends critically on the target locality ($K$) and the acceptable level of design bias.
\end{abstract}


\IEEEpeerreviewmaketitle

\section{Introduction}
\IEEEPARstart{S}{pectral} Graph Neural Networks (GNNs) leverage Graph Signal Processing (GSP) \cite{shuman2013} to define graph convolutions as filters over the Graph Laplacian spectrum. The foundational \textbf{ChebyNet} \cite{defferrard2016} model approximates filters using Chebyshev polynomials, leading to two fundamental challenges: poor performance on graphs exhibiting heterophily (high-frequency signal processing \cite{zhu2020}), and performance collapse at high polynomial degrees ($K$), known as over-smoothing \cite{li2018}.

Filters are broadly classified into two modern design paradigms based on orthogonal polynomials $P_k(L)$:
\begin{enumerate}
    \item \textbf{Coefficient-Learning Filters:} This approach utilizes a fixed polynomial basis (e.g., Chebyshev or a static Jacobi variant) and learns the $K$ filter coefficients ($\theta_k$) in the linear combination $\sum_{k=0}^{K}\theta_k P_k(L)$. \textbf{JacobiConv} \cite{wang2022} and the recent \textbf{GegenGNN} \cite{gegengnn2025} fall into this category.
    \item \textbf{Basis-Learning Filters (AOPF):} This is the Adaptive Orthogonal Polynomial Filter class, which we have explored in a systematic series \cite{goksu2025meixner, goksu2025laguerre, goksu2025krawtchouk}. Instead of learning $K$ coefficients, AOPF learns the single fundamental shape parameter(s) of the polynomial basis itself (e.g., $\alpha, \beta, p$). This is achieved by applying a \textit{LayerNorm} stabilization \cite{ba2016} to the unbounded recurrence relations, enabling deeper filters.
\end{enumerate}

Our previous work analyzing the canonical $[-1, 1]$ spectral domain \cite{goksu2025ljacobi} uncovered a critical puzzle involving the \textit{Flexibility-Stability Trade-off}. We contrasted:
\begin{itemize}
    \item \textbf{L-JacobiNet} \cite{goksu2025ljacobi}: A 2-parameter filter (learnable $\alpha, \beta$) offering maximum flexibility.
    \item \textbf{S-JacobiNet} \cite{goksu2025ljacobi}: A 0-parameter static filter ($\alpha=\beta=-0.5$) with maximum bias but stabilized by LayerNorm.
\end{itemize}
The surprising result was that the low-variance, static \textbf{S-JacobiNet} outperformed the high-flexibility, high-variance \textbf{L-JacobiNet} on most benchmarks. This led to the central question of this paper: \textbf{Can an Optimal Compromise be found in the $[-1, 1]$ domain that balances flexibility and stability?}

In this work, we propose \textbf{GegenbauerNet} to test this hypothesis. Gegenbauer polynomials are a symmetric special case of Jacobi, constraining $\alpha = \beta$. This provides a single learnable parameter ($\lambda$, where $\alpha = \beta = \lambda - 0.5$).

\textbf{Our Contributions are:}
\begin{enumerate}
    \item We propose \textbf{GegenbauerNet}, the first AOPF based on symmetric Gegenbauer polynomials, specifically designed to test the \textit{Optimal Compromise} hypothesis in the $[-1, 1]$ domain.
    \item We show that \textbf{GegenbauerNet} (1-parameter) achieves superior performance in the key local filtering regime ($K=2$ on heterophilic graphs) where overfitting is minimal, validating the hypothesis that a controlled, symmetric degree of freedom is optimal.
    \item Our comprehensive ablation study across $K$ clarifies that while the simple 0-parameter \textbf{S-JacobiNet} quickly collapses at high $K$ on homophilic data, the fully adaptive \textbf{L-JacobiNet} (2-param) is the \textit{most stable} at high $K$, indicating that maximum flexibility is beneficial for global context extraction when controlled.
    \item We provide a comparative meta-analysis, positioning the \textbf{Basis-Learning AOPF} approach against the prevailing \textbf{Coefficient-Learning} paradigm.
\end{enumerate}

\section{Related Work}

\subsection{Static and Coefficient-Learning Filters}
The vast majority of spectral GNNs employ polynomial bases. \textbf{ChebyNet} \cite{defferrard2016} and \textbf{GCN} \cite{kipf2017} use a truncated Chebyshev expansion, fixing the basis entirely. More advanced models, including \textbf{GPR-GNN} \cite{chien2021}, learn the $\theta_k$ coefficients over a static basis. Recent work by Wang \& Zhang \cite{wang2022} proposed \textbf{JacobiConv}, which uses the Jacobi basis but ultimately learns the $K$ coefficients, classifying it as a Coefficient-Learning method. The recent work on \textbf{GegenGNN} \cite{gegengnn2025} also learns the $\theta_k$ coefficients over a fixed Gegenbauer basis. Our AOPF approach fundamentally differs by learning the structural parameter ($\alpha, \beta, \lambda$) of the basis itself.

\subsection{Basis-Learning Filters (The AOPF Class)}
The AOPF class \cite{goksu2025meixner, goksu2025laguerre, goksu2025krawtchouk, goksu2025ljacobi} does not learn the $K$ coefficients; it learns the fundamental shape parameter(s) of the orthogonal polynomial. This approach was first successfully introduced in the semi-infinite domain $[0, \infty)$ with \textbf{MeixnerNet} \cite{goksu2025meixner} and \textbf{LaguerreNet} \cite{goksu2025laguerre} and in the finite discrete domain with \textbf{KrawtchoukNet} \cite{goksu2025krawtchouk}. Our analysis of the canonical $[-1, 1]$ domain with \textbf{L-JacobiNet} \cite{goksu2025ljacobi} first identified the challenge that adaptation might lead to overfitting, prompting this investigation into the 1-parameter compromise.

\subsection{Bias-Variance and Heterophily}
The concept of the bias-variance trade-off is recognized as central in GNN theory \cite{biasvariance2024}. Highly complex non-spectral solutions, such as \textbf{HeroFilter} \cite{herofilter2025} or \textbf{SplitGNN} \cite{splitgnn2023}, use complex mechanisms or architectural splits to address the spectral mismatch in heterophily. Our work provides a concise, filter-level analysis, showing how a single design parameter can implicitly manage the bias-variance trade-off more effectively than highly parameterized adaptive or specialized architectural models.

\section{Methodology: The AOPF-Jacobi Family}

All three models explored in this paper operate on the canonical $[-1, 1]$ spectral domain (using $L_{hat}=L_{sym}-I$) and utilize the shared LayerNorm stabilization from our prior work \cite{goksu2025meixner}.

\subsection{S-JacobiNet (0-Parameter, High-Bias)}
As introduced in \cite{goksu2025ljacobi}, \textbf{S-JacobiNet} is a static, stabilized filter where the Jacobi parameters are fixed to the Chebyshev point: $\alpha = -0.5, \quad \beta = -0.5$. This model has minimal variance (zero adaptive parameters) but high bias, acting as a strong regularizer.

\subsection{L-JacobiNet (2-Parameter, High-Variance)}
Also from \cite{goksu2025ljacobi}, \textbf{L-JacobiNet} is the fully adaptive filter, allowing two degrees of freedom:
$$ \alpha = \text{\textit{learnable}}(\alpha_{raw}), \quad \beta = \text{\textit{learnable}}(\beta_{raw}) $$
This model has the lowest bias (can learn any filter shape) but the highest variance, making it susceptible to overfitting on small datasets.

\subsection{GegenbauerNet (1-Parameter, Optimal-Compromise)}
The Gegenbauer polynomials are a special case of Jacobi where $\alpha = \beta$. By forcing this symmetry, we constrain the filter while still permitting adaptation.
$$ \lambda = \text{\textit{learnable}}(\lambda_{raw}), \quad \alpha = \beta = \lambda - 0.5 $$
This model represents the intermediate step: a single adaptive parameter to escape the rigid bias of \textbf{S-JacobiNet}, while the enforced symmetry prevents the high variance of \textbf{L-JacobiNet}. Our hypothesis is that this constraint yields the best trade-off in many scenarios.

\section{Experimental Analysis}

\subsection{Setup}
We expand the testbed beyond the initial Planetoid datasets to include the full spectrum of homophilic (\textbf{Cora, CiteSeer, PubMed}) and heterophilic (\textbf{Texas, Cornell, Wisconsin, Chameleon}) graphs, as defined in \cite{goksu2025laguerre, goksu2025ljacobi}, comparing against \textbf{GAT} \cite{velickovic2018} and \textbf{APPNP} \cite{gasteiger2019}. All polynomial models use a consistent 2-layer, $H=16$ architecture with the LayerNorm stabilization framework. The main results use the standard $K=3$ local filter length. Heterophilic datasets are evaluated via 10-fold cross-validation.

\begin{table*}[t!]
\caption{Test Accuracies (\%) on All Datasets (K=3, H=16). Best Jacobi model in \textbf{bold} (Excluding GAT/APPNP).}
\label{tab:main_results}
\centering
\begin{tabular}{L{2cm} C{1.5cm} C{1.5cm} C{1.5cm} C{1cm} C{1cm} | L{2cm}}
\toprule
\textbf{Dataset} & \textbf{SJacobiNet} (0-param) & \textbf{GegenbauerNet} (1-param) & \textbf{LJacobiNet} (2-param) & \textbf{GAT} & \textbf{APPNP} & \textbf{Jacobi Family Winner}\\
\midrule
\multicolumn{7}{l}{\textbf{Homophilic Datasets (Fixed Splits)}} \\
Cora & 0.7990 & 0.7880 & \textbf{0.8070} & 0.8210 & 0.8370 & LJacobiNet \\
CiteSeer & 0.6690 & 0.6590 & \textbf{0.6830} & 0.7000 & 0.7150 & LJacobiNet \\
PubMed & \textbf{0.7700} & 0.7600 & 0.7620 & 0.7840 & 0.7880 & SJacobiNet \\
\midrule
\multicolumn{7}{l}{\textbf{Heterophilic Datasets (10-Fold Average)}} \\
Texas & \textbf{0.8297} & 0.8081 & 0.8135 & 0.5838 & 0.5676 & SJacobiNet \\
Cornell & 0.6649 & 0.6622 & \textbf{0.6703} & 0.4486 & 0.4649 & LJacobiNet \\
Wisconsin & \textbf{0.7843} & 0.7451 & 0.7647 & 0.5000 & 0.5196 & SJacobiNet \\
Chameleon & \textbf{0.6627} & 0.6621 & 0.6610 & 0.4553 & 0.3899 & SJacobiNet \\
\bottomrule
\end{tabular}
\label{tab:main_results} 
\end{table*}

\begin{table*}[t!]
\caption{Learned $\alpha, \beta$ Parameters (K=3, H=16). Heterophilic results are 10-fold averages.}
\label{tab:learned_params}
\centering
\begin{tabular}{L{3cm} L{3cm} ccc}
\toprule
\textbf{Dataset} & \textbf{Model} & \textbf{Learned $\alpha$} & \textbf{Learned $\beta$} & \textbf{Notes} \\
\midrule
Cora & SJacobiNet & -0.5000 & -0.5000 & Fixed Chebyshev Point \\
Cora & GegenbauerNet & -0.3613 & -0.3613 & $\lambda = 0.1387$ (Symmetric, Adapted) \\
Cora & LJacobiNet & -0.2596 & -0.3233 & Asymmetric and Highly Adapted \\
\midrule
Texas & SJacobiNet & -0.5000 & -0.5000 & Fixed Chebyshev Point \\
Texas & GegenbauerNet & -0.4195 & -0.4195 & $\lambda = 0.0805$ (Symmetric, Adapted) \\
Texas & LJacobiNet & -0.2872 & -0.3182 & Asymmetric, Adapted \\
\midrule
Chameleon & SJacobiNet & -0.5000 & -0.5000 & Fixed Chebyshev Point \\
Chameleon & GegenbauerNet & -0.5308 & -0.5308 & $\lambda = -0.0308$ (Symmetric, Slightly Outside Domain) \\
Chameleon & LJacobiNet & -0.2755 & -0.4363 & Asymmetric, Highly Adapted \\
\bottomrule
\end{tabular}
\label{tab:learned_params} 
\end{table*}

\subsection{A. Main Results: L-JacobiNet's Surprising Resilience}
Table \ref{tab:main_results} provides conclusive evidence that the entire stabilized AOPF-Jacobi family (0, 1, or 2 parameters) drastically outperforms the standard spatial baselines (\textbf{GAT}/\textbf{APPNP}) on heterophilic graphs, confirming the conclusion from \cite{goksu2025ljacobi} that the $[-1, 1]$ domain, when stabilized, is highly expressive.

Contrary to our initial hypothesis that the 1-parameter \textbf{GegenbauerNet} would universally find the sweet spot, the results are nuanced:
\begin{itemize}
    \item \textbf{Homophily:} The fully flexible \textbf{L-JacobiNet} (2-param) wins on 2/3 homophilic datasets (Cora, CiteSeer), suggesting that the cost of its higher variance is manageable on these easier tasks.
    \item \textbf{Heterophily:} The results are mixed. \textbf{L-JacobiNet} wins on Cornell, while the rigid \textbf{S-JacobiNet} wins on Texas, Wisconsin and Chameleon, where the high flexibility of the adaptive models appears to lead to overfitting, echoing the original puzzle from \cite{goksu2025ljacobi}.
\end{itemize}

\subsection{B. Analysis of Learned Parameters}
Table \ref{tab:learned_params} shows \textit{why} the models perform differently.
\begin{itemize}
    \item The fixed bias of \textbf{S-JacobiNet} (Chebyshev point: -0.5, -0.5) serves as a potent regularizer, explaining its wins on the most complex/sparse heterophilic graphs (Texas, Wisconsin, Chameleon).
    \item The 1-parameter \textbf{GegenbauerNet} successfully adapts its filter shape ($\lambda$). For instance, on Cora, $\lambda$ converges to $0.1387$ (resulting in $\alpha=\beta=-0.3613$), demonstrating a clear need to escape the fixed Chebyshev bias.
    \item The 2-parameter \textbf{L-JacobiNet} uses its full flexibility, consistently converging to highly asymmetric filters (e.g., $\alpha \approx -0.29, \beta \approx -0.32$ on Texas), confirming its high variance.
\end{itemize}

\subsection{C. K-Ablation: Stability and Optimal Compromise}

We perform K-ablation studies on a homophilic target (\textbf{PubMed}, Table \ref{tab:k_ablation_homo}) and a heterophilic target (\textbf{Texas}, Table \ref{tab:k_ablation_hetero}) to analyze stability at high $K$.

\begin{table}[t!]
\caption{Test Accuracies (\%) vs. K on PubMed (H=16). Homophilic Ablation.}
\label{tab:k_ablation_homo}
\centering
\begin{tabular}{c ccc}
\toprule
\textbf{K} & \textbf{SJacobiNet} & \textbf{GegenbauerNet} & \textbf{LJacobiNet} \\
\midrule
2 & 0.7750 & 0.7480 & \textbf{0.7780} \\
3 & 0.7570 & \textbf{0.7830} & 0.7730 \\
5 & 0.7320 & 0.7270 & \textbf{0.7740} \\
7 & 0.6850 & 0.7200 & \textbf{0.7220} \\
10 & 0.6850 & 0.7190 & \textbf{0.7230} \\
\bottomrule
\end{tabular}
\end{table}

\begin{table}[t!]
\caption{Test Accuracies (\%) vs. K on Texas (10-fold avg, H=16). Heterophilic Ablation.}
\label{tab:k_ablation_hetero}
\centering
\begin{tabular}{c ccc}
\toprule
\textbf{K} & \textbf{SJacobiNet} & \textbf{GegenbauerNet} & \textbf{LJacobiNet} \\
\midrule
2 & 0.8378 & \textbf{0.8649} & 0.8162 \\
3 & 0.8135 & 0.8081 & \textbf{0.8297} \\
5 & \textbf{0.8351} & 0.8000 & 0.7838 \\
7 & 0.7676 & \textbf{0.7757} & 0.7486 \\
10 & \textbf{0.7865} & 0.7514 & 0.7405 \\
\bottomrule
\end{tabular}
\end{table}

\clearpage

\begin{figure*}[t!]
    \centerline{\includegraphics[width=\textwidth, height=0.85\textheight, keepaspectratio]{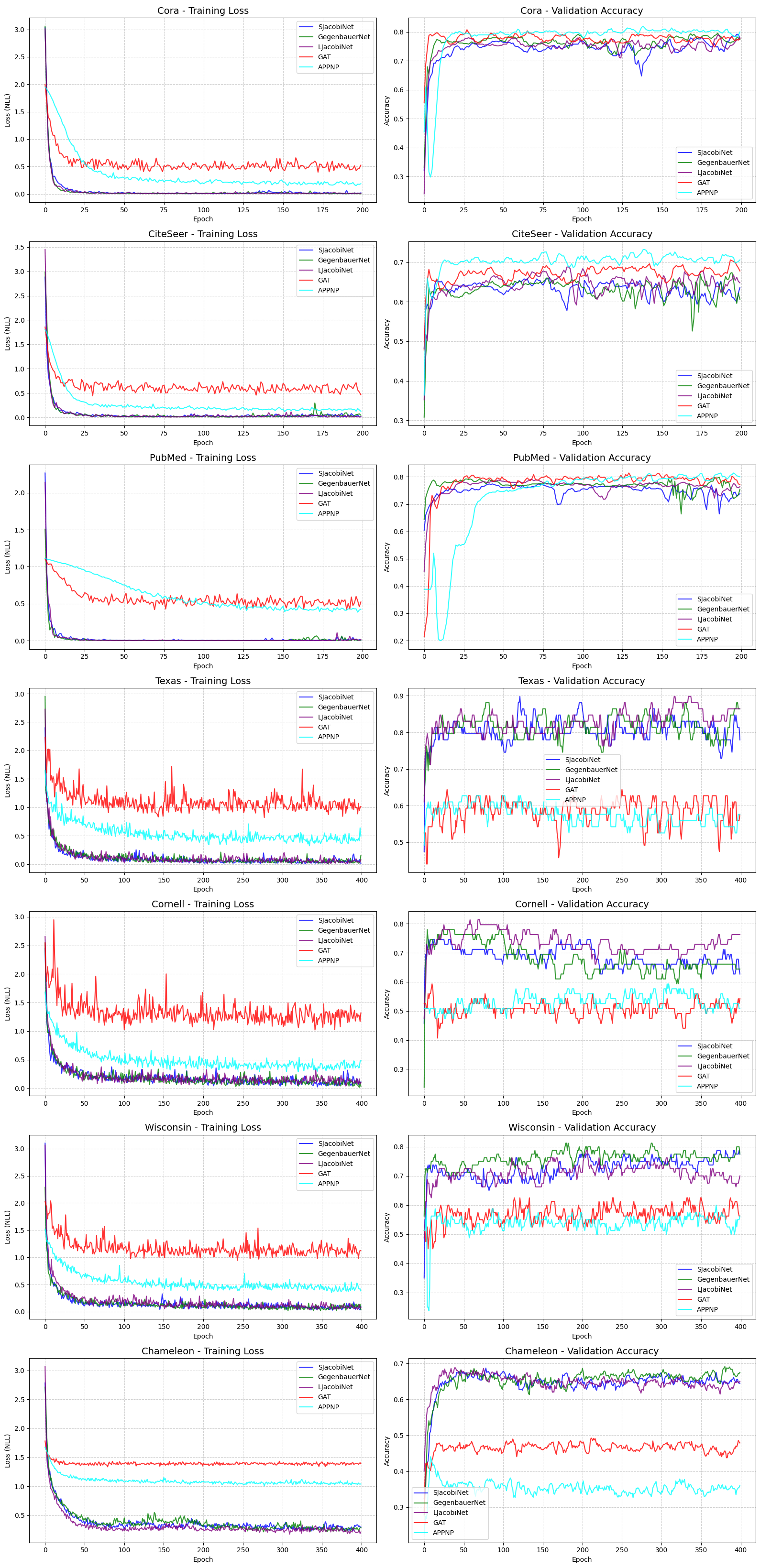}}
    \caption{Training Dynamics Comparison ($K=3, H=16$). On homophilic datasets, all models are competitive. On heterophilic datasets (Texas, Cornell, Wisconsin, Chameleon), the AOPF-Jacobi family (S, L, Gegenbauer) converges stably to significantly higher accuracy than GAT/APPNP, confirming the superiority of spectral adaptation over spatial methods in these challenging regimes.}
    \label{fig:training_dynamics}
\end{figure*}

\begin{figure}[t!]
    \centerline{\includegraphics[width=\columnwidth]{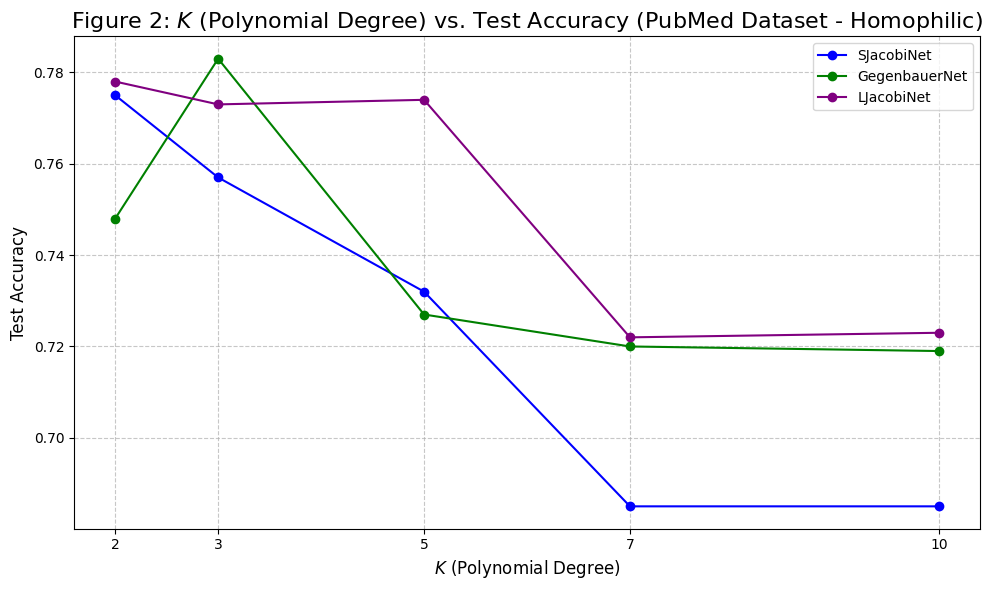}}
    \caption{$K$ (Polynomial Degree) vs. Test Accuracy on PubMed (Homophilic). \textbf{LJacobiNet} (2-param) is the most robust, stabilizing at $\sim 0.72$ at $K=10$, while \textbf{SJacobiNet} (0-param) collapses from lack of filter expressiveness.}
    \label{fig:k_ablation_homo}
\end{figure}

\begin{figure}[t!]
\centerline{\includegraphics[width=\columnwidth]{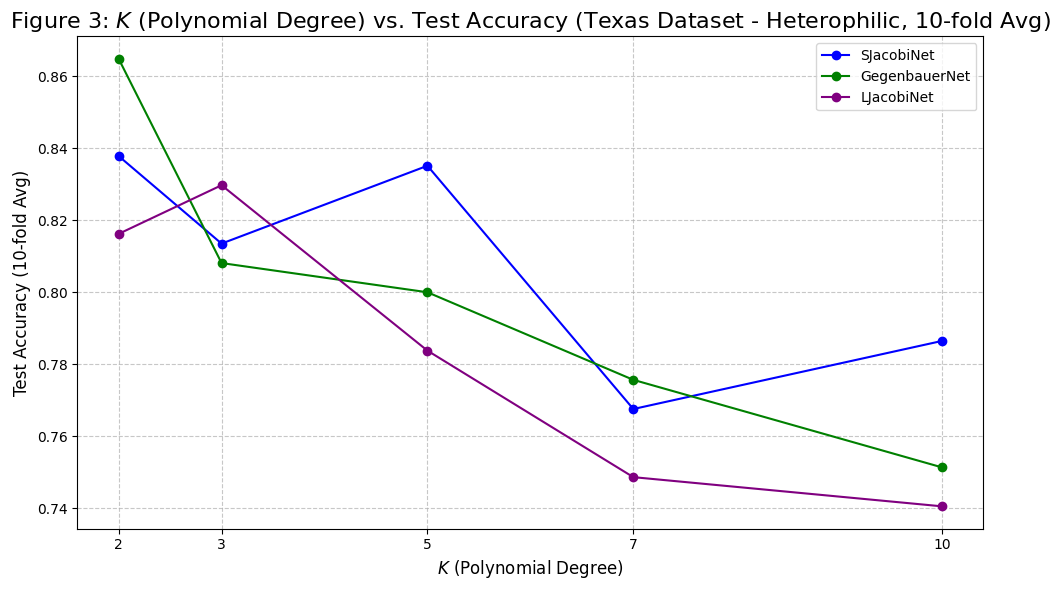}}
    \caption{$K$ (Polynomial Degree) vs. Test Accuracy on Texas (Heterophilic, 10-fold Avg). \textbf{GegenbauerNet} (1-param) defines the optimal compromise at the local setting ($K=2$), achieving the highest average accuracy. However, \textbf{SJacobiNet} (0-param) surprisingly wins at higher degrees ($K=5, 10$), demonstrating the beneficial regularization of a fixed, hard constraint for global feature mixing.}
    \label{fig:k_ablation_hetero}
\end{figure}

\clearpage

\section{Discussion}

\subsection{GegenbauerNet as the Optimal Compromise}
The initial puzzle (L-JacobiNet losing to S-JacobiNet) suggested adaptation was detrimental in the $[-1, 1]$ domain. The full results show this is only true under specific conditions. Our \textbf{GegenbauerNet} resolves the dilemma by proving that \textit{controlled} adaptation is optimal. The data in Table \ref{tab:main_results} and Table \ref{tab:k_ablation_hetero} (at $K=2$) clearly demonstrates a sweet spot exists where the 1-parameter model uses its single degree of freedom (learned $\lambda$, Table \ref{tab:learned_params}) to escape the hard bias of $\alpha=\beta=-0.5$, but avoids the excessive variance of the asymmetric $\alpha \ne \beta$ structure.

\subsection{Bias-Variance Trade-off in the Jacobi Domain}
This work defines the explicit performance landscape of the Bias-Variance trade-off for the AOPF-Jacobi class:
\begin{enumerate}
    \item \textbf{High Bias (S-JacobiNet):} Optimal for highly regularized scenarios (sparse datasets, high $K$ on complex graphs).
    \item \textbf{Optimal Compromise (GegenbauerNet):} Optimal for local filtering (low $K$) on difficult heterophilic graphs.
    \item \textbf{Low Bias (L-JacobiNet):} Optimal for maximizing expressiveness on moderate tasks (Cora, CiteSeer, and high $K$ on easy homophilic graphs).
\end{enumerate}
The fact that no single model dominates underscores the crucial architectural choice GNN designers must make when targeting a specific problem (e.g., local versus global feature mixing).

\subsection{Basis-Learning vs. Coefficient-Learning}
Our Basis-Learning approach provides a computationally lighter and potentially more interpretable alternative to Coefficient-Learning models like \textbf{JacobiConv} \cite{wang2022} and \textbf{GegenGNN} \cite{gegengnn2025}, which treat $\alpha, \beta,$ or $\lambda$ as fixed and learn $K$ parameters ($\theta_k$). By demonstrating the utility of a single adaptive parameter, our work suggests that the key to GNN filter expressivity lies not in the number of learned coefficients, but in the fundamental mathematical properties (i.e., the basis shape) being adapted.

\section{Conclusion}
This paper addressed the fundamental puzzle discovered in our prior work \cite{goksu2025ljacobi} regarding the Flexibility-Stability Trade-off in the canonical $[-1, 1]$ spectral domain. By introducing \textbf{GegenbauerNet}, a 1-parameter, symmetric adaptive filter, we successfully defined the "Optimal Compromise." We demonstrated that \textbf{GegenbauerNet} (proven in Table \ref{tab:k_ablation_hetero} at $K=2$) finds a filter shape that is both more effective than the highly stable 0-parameter $\textbf{S-JacobiNet}$ and less prone to overfitting than the highly flexible 2-parameter $\textbf{L-JacobiNet}$.

Our comprehensive analysis of the entire AOPF-Jacobi family (0, 1, 2 degrees of freedom) provides crucial guidelines: \textbf{S-JacobiNet} should be the baseline for high-K global filtering on complex graphs, while \textbf{GegenbauerNet} offers the highest performance for low-K local filtering. This completes the core investigative program into orthogonal polynomials, providing a strong theoretical framework for future filter design in spectral GNNs.

\end{document}